\documentclass{article}
\usepackage[utf8]{inputenc}

\usepackage[pdfusetitle,colorlinks,plainpages=false]{hyperref}
\usepackage{graphicx}
\usepackage{amssymb}
\usepackage{amsmath}
\usepackage{amsthm}
\usepackage{relsize}
\usepackage{xcolor}
\usepackage{listings}
\usepackage{hyperref}
\usepackage{subcaption}
\usepackage{mwe}
\usepackage{ulem}
\usepackage{authblk}

\newcommand{\bigsum}[3]{\mathlarger{\mathlarger{\sum}}_{#2}^{#1}{#3}}

\newcommand{\alphaDual}[0]{\overline{\alpha}}
\newcommand{\betaDual}[0]{\overline{\beta}}

\newcommand{\Pw}[1]{$\cal{P}$$(#1)$}

\newcommand{\patrickv}[2]{#2}

\begin{document}

\title{A computation of D(9) using FPGA Supercomputing}
\author[1,2,3]{Lennart Van Hirtum}
\author[1]{Patrick De Causmaecker}
\author[1]{Jens Goemaere}
\author[2,3]{Tobias Kenter}
\author[2,3]{Heinrich Riebler}
\author[2,3]{Michael Lass}
\author[2,3]{Christian Plessl}
\affil[1]{KU Leuven, Department of Computer Science, KULAK}
\affil[2]{Department of Computer Science, Paderborn University}
\affil[3]{Paderborn Center for Parallel Computing, Paderborn University}
\affil[ ]{e-mail adresses in footnote\footnote{lennart.vanhirtum@gmail.com, patrick.decausmaecker@kuleuven.be, jens.goemaere@kuleuven.be, kenter@uni-paderborn.de, heinrich.riebler@uni-paderborn.de, michael.lass@uni-paderborn.de, christian.plessl@uni-paderborn.de}}
\date{April 2023}

\maketitle

\begin{abstract}
    This preprint makes the claim of having computed the $9^{th}$ Dedekind Number. This was done by building an efficient FPGA Accelerator for the core operation of the process, and parallelizing it on the Noctua 2 Supercluster at Paderborn University. The resulting value is $$286386577668298411128469151667598498812366$$ This value can be verified in two steps. We have made the data file containing the 490M results available, each of which can be verified separately on CPU, and the whole file sums to our proposed value. 
\end{abstract}

\section{Introduction}

Let us consider the finite set $A = \{1,\dots,n\}$, which we will call the {\it base set}, and let us denote the subsets of $A$ by \Pw{A}.
Dedekind numbers count the number of monotone Boolean functions on \Pw{A}.
This number is denoted by $D(n)$ and it is called {\it the $n^{th}$ Dedekind number}.
%Unless explicit reference to the elements of $A$ is needed, we will denote the set of monotone Boolean functions on \Pw{A}, with $|A|= n$, by $D_n$, otherwise we will use $D_n$.

The set of permutations of the elements of base set $A$ generates an equivalence relation on $D_n$.
The equivalence classes of this relation are denoted by $R_n$ and the number of such equivalence classes is denoted by $R(n)$.

Richard Dedekind first defined the numbers $D(n)$ in 1897 \cite{Dedekind1897}.
Over the previous century, Dedekind numbers have been a challenge for computational power in the evolving domain of computer science.
Computing the numbers proved exceptionally hard, and so far only formula's with a double exponential time complexity are known.
Until recently, the largest known Dedekind number was $D(8)$. 
In this paper, we report on a first result for $D(9)$.
Table \ref{tab:dedekindNumbers} shows the known numbers, including the first result of our computation.
As we explain below, a verification run is needed.
We expect to have verified our result in about three months time.

Table \ref{tab:equivalenceClassCounts} shows the known numbers $R(n)$ of equivalence classes of monotone Boolean functions under permutation of the elements of the base set.
Note that the last result dates from 2021.

\begin{table}[h]
    \centering
    \begin{tabular}{c|l|c}
        D(0) & 2 & Dedekind (1897) \\
        D(1) & 3 & Dedekind (1897) \\
        D(2) & 6 & Dedekind (1897) \\
        D(3) & 20 & Dedekind (1897)\\
        D(4) & 168 & Dedekind (1897) \\
        D(5) & 7581 & Church (1940) \\
        D(6) & 7828354 & Ward (1946) \\
        D(7) & 2414682040998 & Church (1965) \\
        D(8) & 56130437228687557907788 & Wiedemann (1991) \\
        D(9) & 286386577668298411128469151667598498812366 & Our Proposal (2023)
    \end{tabular}
    \caption{Known Dedekind Numbers \cite{wiedemannDedekind8} and our first result.}
    \label{tab:dedekindNumbers}
\end{table}

\begin{table}[h]
    \centering
    \begin{tabular}{c|l|c}
        R(0) & 2 & \\
        R(1) & 3 & \\
        R(2) & 5 & \\
        R(3) & 10 & \\
        R(4) & 30 & \\
        R(5) & 210 & \\
        R(6) & 16353 & \\
        R(7) & 490013148 & Tamon Stephen \& Timothy Yusun (2014) \cite{STEPHEN201415} \\
        R(8) & 1392195548889993358 & Bartłomiej Pawelski (2021) \cite{https://doi.org/10.48550/arxiv.2108.13997}
    \end{tabular}
    \caption{Known Equivalence Class Counts}
    \label{tab:equivalenceClassCounts}
\end{table}

Note that a monotone Boolean function is completely defined by the set of sets which are maximal among the sets for which the function value is true.
For any monotone Boolean function, no two of its maximal sets include each other.
This set of sets is called an anti-chain.
Since a monotone Boolean function is completely determined by its associated anti-chain, and any anti-chain is completely determined by its associated monotone Boolean function, we will use any of the two representations whenever it is more convenient. 
We will represent monotone Boolean functions or anti-chains by letters from the Greek alphabet.
If we say that $X \in \alpha$, we mean that $X$ is a maximal set among the sets for which $\alpha$ is $True$, in other words 
$$\forall Y \subseteq X:\alpha(Y) = True\ and\ \forall Z \supsetneq X: \alpha(Z) = False$$
If we say that $\alpha = \{X,Y,Z\}$, we mean that the sets $X,Y,Z \subseteq A$ are the maximal sets among the sets for which $\alpha$ is $True$.
For the set $D_n$ of monotone Boolean functions on the base set a natural partial order $\le$ is defined by
\begin{equation}
    \forall \alpha,\beta \in D_n: \alpha \le \beta \Leftrightarrow \forall X \subseteq A: \alpha(X) \Rightarrow \beta(X)
\end{equation}
This partial ordering defines a complete lattice on $D_n$.
We denote by $\bot$ and $\top$ the smallest, respectively the largest, element of $D_n$:
\begin{align}
    \forall X \subseteq A: \bot(X) = False, \top(X) = True\\
    \bot(X) = \{\}, \top(X) = \{A\}
\end{align}
Intervals in $D_n$ are denoted by
\begin{equation}
    \forall \alpha, \beta \in D_n:[\alpha,\beta] = \chi \in D_n:\alpha \le \chi \le \beta
\end{equation}
For $\alpha,\beta \in D_n$, the {\it join} $\alpha \vee \beta$ and the {\it meet} $\alpha \wedge \beta$ are the monotone Boolean functions defined by

\begin{align}
    \forall X \subseteq A & :(\alpha \vee \beta)(X) =  \alpha(X)\ or\ \beta(X)\\
    \forall X \subseteq A & :(\alpha \wedge \beta)(X) =  \alpha(X)\ and\ \beta(X)
\end{align}

Finally, in the formulas below, a number defined for each pair $\alpha \le \beta \in D_n$ plays an important role.
We refer to this number as the {\it connector number}  $C_{\alpha,\beta}$ of $\alpha$ and $\beta$.
It counts the number of connected components of the anti-chain $\beta$ with respect to $\alpha$.
Two such sets $X, Y \in \beta$ are connected if $\alpha(X \cap Y) = False$ or if there is a path $X,Z_1,...,Z_n,Y$ of such subsets $X,Z_1,...,Z_n,Y \subseteq A$ in which for every two subsequent sets $\alpha(X \cap Z_1) = \alpha(Z_1 \cap Z_2) = ... = \alpha(Z_n \cap Y) = False$.
It turns out that the number of solutions of 
\begin{align}
    \chi \vee \upsilon = \beta\\
    \chi \wedge \upsilon = \alpha
\end{align}
for $\chi,\upsilon \in D_n$ is given by $2^{C_{\alpha,\beta}}$. This is called the {\it PCoeff} \cite{decausmaecker2014number,decausmaecker2016intervals}.

\section{Method, Theory}
The original PCoeff Formula as taken from \cite{decausmaecker2014number}. 
\begin{equation}
    D(n+2) = \bigsum{}{\alpha, \beta \in D_n}{|[\bot,\alpha]|2^{C_{\alpha,\beta}}|[\beta,\top]|}
\label{eq:pcoeffEq}
\end{equation}

\patrickv{In my thesis}{In the master thesis of the first author of the current paper, Lennart Van Hirtum} \cite{lennartThesis}\patrickv{I reworked}{, the author reworked} this formula to a form making use of equivalence classes to reduce the total number of terms. 
\begin{equation}
    D(n+2) = \bigsum{}{\alpha \in R_n}{|[\bot,\alpha]|D_\alpha\bigsum{}{\substack{\beta \in R_n \\ \exists \delta \simeq \beta : \alpha \leq \delta}}{|[\beta, \top]|\frac{D_\beta}{n!}\bigsum{}{\substack{\gamma \in Permut_\beta \\ \alpha \leq \gamma}}{2^{C_{\alpha,\gamma}}}}}
\label{eq:eqClassIteration}
\end{equation}
The $Permut_\beta$ term is the collection of all $n!$ equivalents of $\beta$ under permutation of the base set. $D_\beta$ is the number of different equivalents, and hence, $Permut_\beta$ contains duplicates iff $D_\beta < n!$. These duplicates are divided out by the $\frac{D_\beta}{n!}$ factor.

For D(9), this means iterating through \patrickv{the space of 7 variables}{$D_7$}. 
That would require iterating over an estimated $4.59*10^{16} \alpha, \beta$ pairs. The total number of P-Coëfficients ($C_{\alpha,\gamma}$) that needed to be computed was $1.148*10^{19}$. However we were able to improve on this further using the process of 'deduplication', where we can halve the total amount of work again, by noticing that pairs of $\alpha,\beta$ give identical results to their dual pair $\betaDual,\alphaDual$. As per Equation \ref{eq:dedup}. This allowed us to halve the total amount of work to $5.574*10^{18}$ P-Coëfficients. \footnote{We made sure not to deduplicate pairs that were their own dual, ie when $\beta=\alphaDual$}

\begin{equation}
|[\bot,\alpha]|2^{C_{\alpha,\beta}}|[\beta,\top]| = |[\alphaDual,\top]|2^{C_{\betaDual,\alphaDual}}|[\bot,\betaDual]|
\label{eq:dedup}
\end{equation}

\section{Computing P-Coëfficients on FPGA}
\begin{figure}[h]
    \centering
    \includegraphics[width=0.8\textwidth]{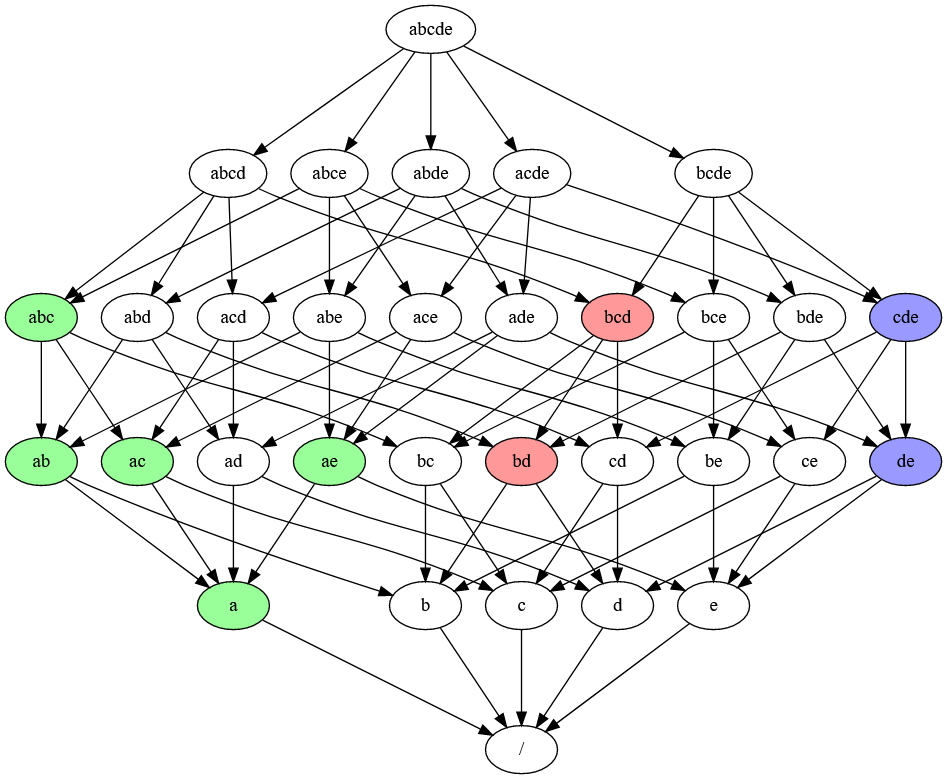}
    \caption[CountConnected Example]{Connected components of an example graph. In this case there are 3 connected components. }
    \label{fig:countConnectedExample}
\end{figure}

\begin{figure}[h]
    \centering
    \includegraphics[width=0.9\textwidth]{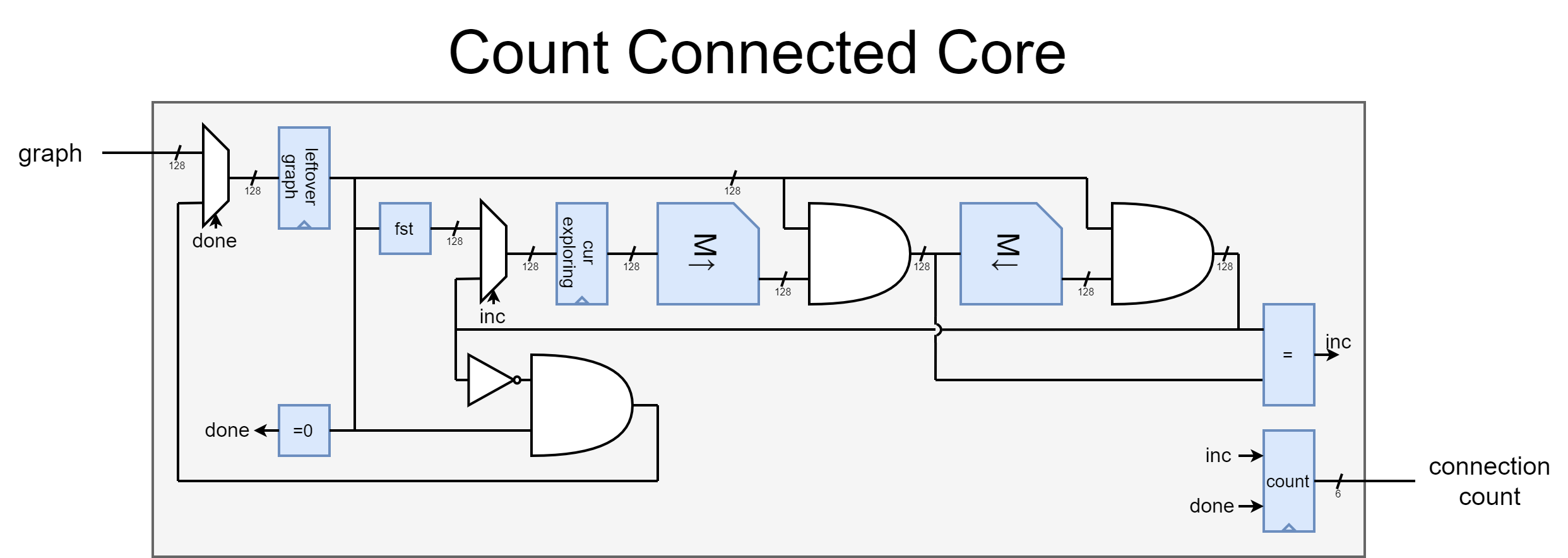}
    \caption{The CountConnected Core}
    \label{fig:countConnectedCore}
\end{figure}

Computing P-Coëfficients is uniquely well-suited for hardware implementation. Computing these terms requires solving the problem of counting the number of distinct connected components within a standard graph structure. An example of such a graph with its distinct connected components colored is shown in Figure \ref{fig:countConnectedExample}. Looking at the connection count problem structure it is easy to see why counting connected components is an incredibly branchy procedure, and why traditional instruction-based computing methods fare poorly on it, especially any kind of Single Instruction, Multiple Data (SIMD) based method. However, since counting connected components is almost purely plain boolean operations, it translates very well to a hardware implementation. A simple schematic implementation is shown in Figure \ref{fig:countConnectedCore}. A detailed explanation of how it works is provided in \patrickv{my Master's Thesis}{the first author's master thesis} \cite{lennartThesis}. \patrickv{In my thesis I show some}{In this thesis, some} optimizations \patrickv{ }{are derived} that bring the average number of iterations down to 4.061. 

\section{Computation on Noctua 2}
We implemented this hardware accelerator on the Intel Stratix 10 GX 2800 cards found in Paderborn University's Noctua 2 supercomputer. We were able to fit 300 of these CountConnected Cores on a single field-programmable gate array (FPGA) die. These CountConnected Cores run at 450MHz. This gives us a throughput of about 33 Billion CountConnected operations per second. At this rate, a single FPGA processes about 5.2 $\alpha$ values per second, taking 47'000 FPGA hours to compute D(9) on Noctua 2, or about 3 months real-time. 

The computation is split across the system along the lines of Equation \ref{eq:eqClassIteration}. $\alpha$ values (also named tops) are divided on the job level. There are 490M tops to be processed for D(9). We split these into 15000 jobs of 30000 tops each. 
The $\beta$ values per top (also named bottoms) are placed in large buffers of 46M bots on average, and sent over PCIe (Peripheral Component Interconnect Express)  to the FPGA. The FPGA then computes all 5040 permutations ($\gamma$) of each bottom, computes and adds up their P-Coëfficients, and stores the result in an output buffer of the same size. 

The artifact of this computation is a dataset with an intermediary result for each of the 490M $\alpha$ values. Each of these can be checked separately\footnote{It takes about 10-200s to compute a single $\alpha$ result on 128 AMD Epyc CPU cores}, and the whole file sums to 286386577668298411128469151667598498812366. 

\begin{figure}[h]
    \centering
    \includegraphics[width=0.3\textwidth]{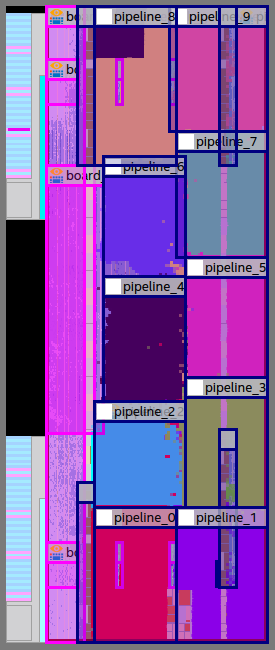}
    \caption{The FPGA Accelerator Die}
    \label{fig:acceleratorDie}
\end{figure}

\section{Correctness}
As much of the code as possible is written generically. This means the same system is used for computing D(3) - D(8). All of these yield the correct results. Of course, the FPGA kernel is written specifically for D(9) computation, so its correctness was verified by comparing its results with the CPU results for a small sample. In effect, both methods verified each other's correctness.

However, we do have several checks to increase our confidence in the result. 

\begin{itemize}
    \item The most direct is the $D(9) \equiv 6\ mod\ 210$ check provided by Pawelski \& Szepietowski\cite{pawelski2023divisibility}. Our result passes this check. Sadly, due to the structure of our computation, nearly all terms are divisible by 210, which strongly hampers the usefulness of this check. One thing that this check does give us is that no integer overflow has occurred, which was an important concern given we were working with integers of 128 and 192 bits wide. 
    \item Our computation was plagued by one issue in particular. Namely that there is a bug in the vendor library for communication over PCIe, wherein, occasionally and at a low incidence rate, full 4K pages of FPGA data are not copied properly from FPGA memory to host memory. This results in large blocks of incorrect bottoms for some tops. We encountered this issue in about 2300 tops. We were able to mitigate this issue by incuding extra data from the FPGA to host memory, namely the 'valid permutation count'. By checking these values, we could determine if a bottom buffer had been corrupted. Additionally, adding all of these counts yields the value for D(8), which shows that the correct number of terms have been added. 
    \item Finally, there is an estimation formula, which gives us an estimation which is relatively close to our result. The Korshunov estimation formula estimates D(9) = $1.15 * 10^{41}$ which is off by about a factor 2. \footnote{This isn't too unusual though, as the results for odd values are off by quite a lot. Estimation for D(3) overestimates by a factor 2, D(5) also overestimates by a factor 2, and D(7) overestimates roughly 10\%}
\end{itemize}

\section{Remaining potential source of errors}
The one way our result could still be wrong is due to a Single Event Upset (SEU), such as a bitflip in the FPGA fabric during processing, or a bitflip during data transfer from FPGA DDR memory to Main Memory.

It is \patrickv{difficult}{hard} to characterise the odds of these SEU events. 
\patrickv{Finding concrete numbers for these FPGAs is almost impossible:}{Expected number of occurrences for the FPGA's we used are not available to the best of our knowledge.} But example values shown on Intel's website pin the error rate at around 5000 SEU events per billion FPGA hours. In that case, given our 47000 FPGA hours, we expect to see 0.235 errors Poisson distributed, giving us a chance of 20\% of a hit. Of course, \patrickv{I don't know how realistic this value is, so}{this is just an example and } the real odds might be higher than that. 

\section{Conclusion}
In conclusion, our method for computing D(9) works, our implementation should theoretically give the correct result. All that remains is: Have any bit errors occurred during this first computation? In any case, we're starting up a second run now, and as it progresses we'll gain more and more confidence in our result. Each subresult is computed a second time, and any values that differ are recomputed a third time as a tiebreaker. 
If we find no errors, then we could have been sure that 9th Dedekind Number was found on the 8th of March, 2023 using the Noctua 2 supercluster at Paderborn University. This value was registered in the corresponding github commit: \url{https://github.com/VonTum/Dedekind/commit/1cf7b019afca655586e8210f97fbb5399d61e842}
All code is available at \url{https://github.com/VonTum/Dedekind}. 

\section{Note}
On April 4th a preprint claiming D(9) was published, right before the present publication. "A computation of the ninth Dedekind Number A Preprint" by Jäkel Christian \cite{ninthFoundRightBeforeMe}. This paper confirms our result. This shows our tentative result is in fact correct. 

\bibliographystyle{unsrt}
\bibliography{refs} 

\end{document}